\documentclass[10pt]{article}
\usepackage[OE]{express}

\newcommand{\Be}{\ensuremath{^9{\rm Be}^+}}

\begin{document}
\title{Third-harmonic-generation of a diode laser for quantum control of beryllium ions}

\author{Ryan A. Carollo, David A. Lane, Edward K. Kleiner, Phyo Aung Kyaw, Chu C. Teng, Celia Y. Ou, Shenglan Qiao and David Hanneke\authormark{*}}

\address{Physics \& Astronomy Department, Amherst College, Amherst, Massachusetts 01002, USA}

\email{\authormark{*}dhanneke@amherst.edu}

\begin{abstract*}
We generate coherent ultraviolet radiation at 313~nm as the third harmonic of an external-cavity diode laser. We use this radiation for laser cooling of trapped beryllium atomic ions and sympathetic cooling of co-trapped beryllium-hydride molecular ions. An LBO crystal in an enhancement cavity generates the second harmonic, and a BBO crystal in a doubly resonant enhancement cavity mixes this second harmonic with the fundamental to produce the third harmonic. Each enhancement cavity is preceded by a tapered amplifier to increase the fundamental light. The 36-mW output power of this all-semiconductor-gain system will enable quantum control of the beryllium ions' motion.
\end{abstract*}


\section{Introduction}
Trapped atomic ions have proven useful in a variety of fundamental and applied physics experiments. The beryllium ion (\Be) in particular has been used for quantum information processing~\cite{winelandJRNIST1998,winelandRMP2013}, quantum simulation~\cite{brittonNature2012}, quantum logic spectroscopy for optical atomic clocks~\cite{schmidtScience2005,rosenbandPRL2007}, as well as sympathetic cooling of molecules~\cite{blythePRL2005,rothJPB2006} and highly charged ions~\cite{schmogerScience2015}. The lowest transition in \Be~from the ground state to an excited electronic state is in the ultraviolet (UV) near 313~nm~\cite{bollingerPRA1985}. There are no convenient materials with optical gain at that wavelength, so \Be~laser systems require nonlinear frequency conversion of one or more laser sources. Such systems balance complexity, cost, and output power. Existing techniques include second-harmonic generation (SHG) of dye lasers~\cite{bollingerPRA1985}, doubly resonant sum frequency generation (SFG) of a Ti:sapphire laser with a frequency-doubled Nd:YAG laser~\cite{schnitzlerAO2002}, fifth-harmonic generation of an amplified fiber laser~\cite{vasilyevAPB2011}, frequency-doubled light from SFG of two amplified fiber lasers~\cite{wilsonAPB2011,loAPB2014}, fourth-harmonic generation of an amplified diode laser, and SHG of diode lasers tuned by extreme temperatures~\cite{cozijnOL2013}. The only published result using these temperature-tuned diode lasers achieved 35~$\mu$W of UV light~\cite{cozijnOL2013}, though outputs in the few-milliwatt range should be possible~\cite{ballRSI2013}. 

\begin{figure}[tb]
\centering
\includegraphics[width=\linewidth]{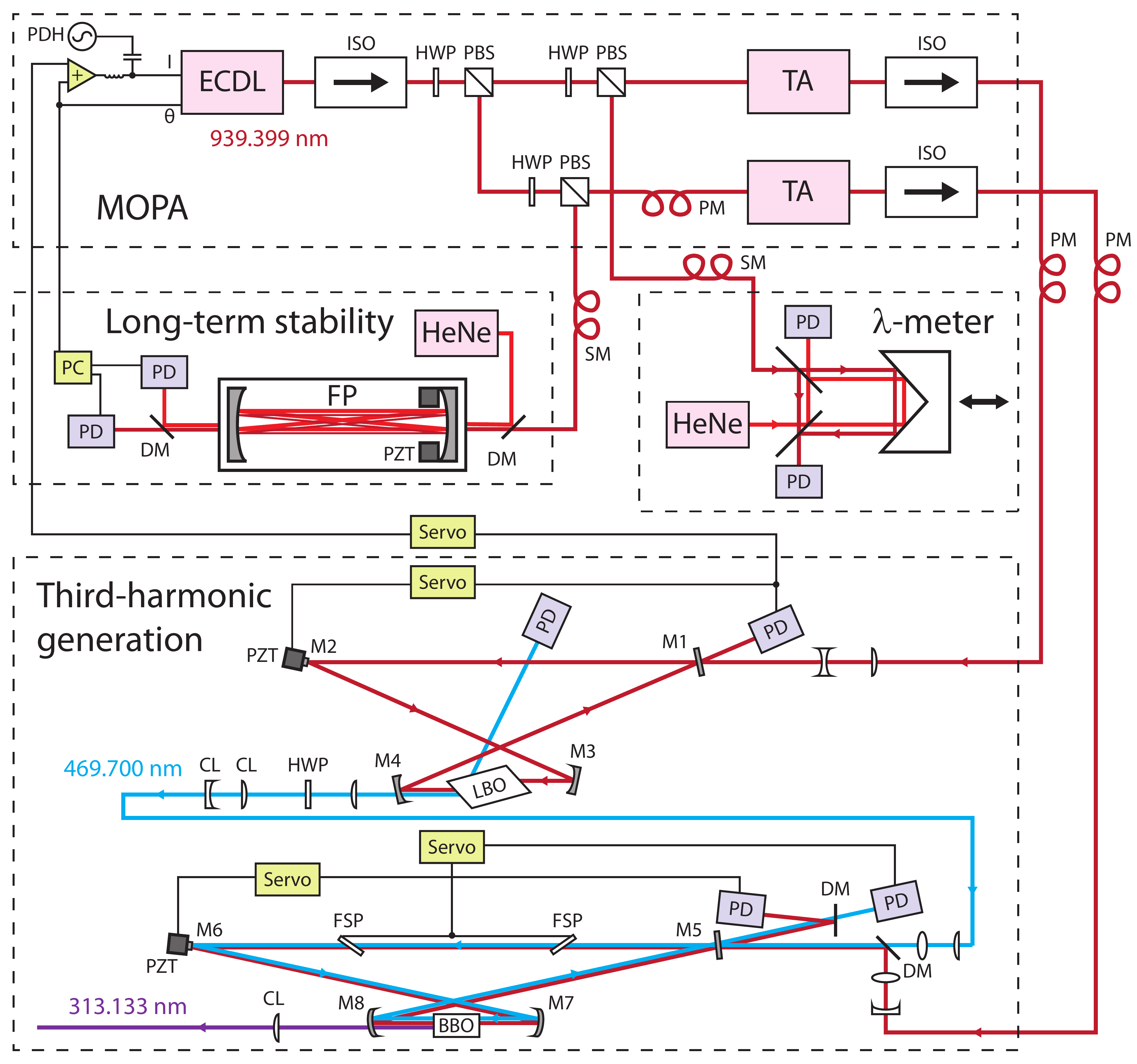}
\caption{Schematic of the laser system. For simplicity, it omits many mirrors, lenses, and diagnostic and control electronics not mentioned in the text. CL: cylindrical lens, DM: dichroic mirror, ECDL: external-cavity diode laser, FP: Fabry--P\'{e}rot cavity, FSP: fused-silica plate, HeNe: stabilized helium-neon laser, HWP: half-waveplate, I/$\theta$: laser current or grating-angle control, ISO: optical isolator, MOPA: master-oscillator power amplifier, PBS: polarizing beamsplitter, PC: computer, PD: photodetector, PDH: Pound--Drever--Hall local oscillator, PM: polarization-maintaining fiber, PZT: piezoelectric transducer, SM: single-mode fiber, TA: tapered amplifier}
\label{fig:beamline}
\end{figure}

Our approach combines the versatility and low cost of a diode laser with the relatively higher powers available with tapered amplifiers in the infrared (IR). By use of two nonlinear crystals in power-enhancement cavities, we generate the third harmonic of 939~nm. Our system's output power is 1000 times that of Ref.~\cite{cozijnOL2013}, though it comes with the added complexity of a doubly resonant SFG stage. Fig.~\ref{fig:beamline} provides a schematic of the system. An external-cavity diode laser (ECDL) serves as a tunable master laser at the fundamental frequency. The first nonlinear stage doubles the frequency in lithium triborate (LBO, LiB$_3$O$_5$).
This second-harmonic light is summed with the fundamental in $\beta$-barium borate (BBO, $\beta$-BaB$_2$O$_4$) 
in a doubly resonant cavity. Our approach is similar to that in Ref.~\cite{mesAPL2003}, except we seed with a diode laser instead of a Ti:Sapphire laser. The ability for rapid frequency modulation through the diode laser current allows us to use the doubling cavity itself to narrow the laser linewidth considerably and to lock the enhancement cavities with a phase-modulation (Pound--Drever--Hall, PDH) scheme~\cite{dreverAPB1983} instead of a polarization (Hansch--Couillaud) scheme~\cite{hanschOC1980}. Our demonstrated output power is sufficient that this laser could be used for stimulated Raman transitions with several-gigahertz detunings~\cite{jostThesis2010,hemmerlingAPB2011}.

\section{Master laser}
The master laser is a home-built ECDL. The single-mode diode (Sheaumann M9-940-0200-D5P) is collimated with an aspheric lens. A diffraction grating (Thorlabs GH13-18U, 1800 lines/mm, efficiency $\approx 30\,\%$) in the Littrow configuration provides feedback. The grating and collimation package are held and aligned with aluminum flexure mounts similar to those in Ref.~\cite{ricciOptComm1995}. The assembly is temperature-stabilized near room temperature. Feed-forward~\cite{nayukiOptRev1998,petridisRSI2001} from the piezo voltage to the diode current by use of an analog circuit allows mode-hop-free tuning over at least 7~GHz. A typical output power of the master laser is 70~mW, which provides sufficient power for seeding two tapered amplifiers as well as monitoring and diagnostics. A bias tee allows radiofrequency modulation of the laser current~\cite{kobayashiIEEEQE1982,wiemanRSI1991}. We apply a signal at 14.74~MHz and $-25~{\rm dBm}$, which is used in stabilizing the length of the power-enhancement cavities.

We tune the laser near resonance with the \Be~transition by use of a traveling Michelson interferometer~\cite{hallAPL1976}. This wavelength meter is referenced to a stabilized HeNe laser and measures the IR frequency to several parts in $10^7$.

We stabilize the laser against long-term drift by use of a scanning Fabry--P\'{e}rot cavity and a stabilized HeNe laser~\cite{barryThesis2013}. The cavity has a confocal design and a 750~MHz free-spectral range. Both the IR laser and the HeNe are incident on the cavity, and we monitor the transmission resonances of each while sweeping the cavity length. Dichroic mirrors combine and separate the two beams. After each sweep, a computer determines the locations of each resonance and feeds back to the ECDL grating and current to keep the IR wavelength stable relative to the HeNe wavelength. We enclose the cavity in a sealed tube to avoid pressure changes to the index of refraction of air~\cite{riedleRSI1994}. One scan---feedback cycle takes approximately 40~ms. With this system, the long-term drift of the IR laser should track that of the stabilized HeNe. We have measured drifts of less than 200~kHz over three hours.

\begin{figure}[tb]
\centering
\includegraphics[width=\linewidth]{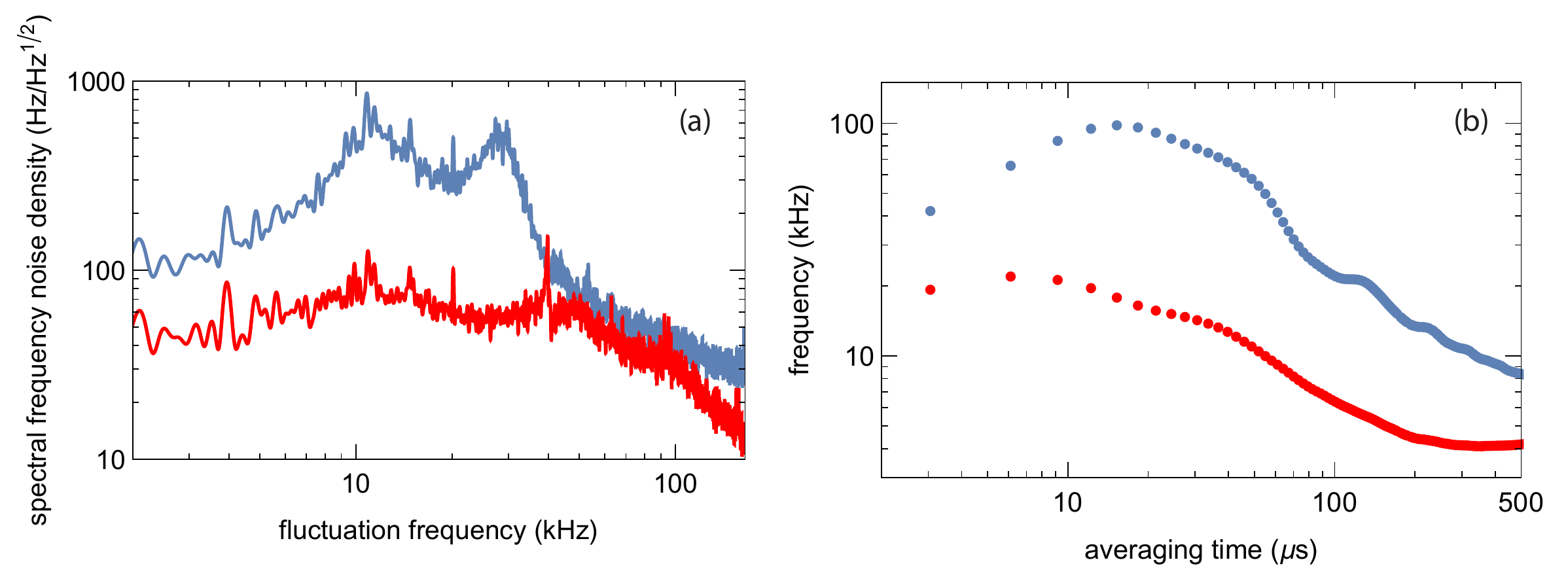}
\caption{Laser frequency noise with (red) and without (blue) current feedback, presented as (a) the spectral frequency noise density and (b) the overlapped Allan deviation.}
\label{fig:noise}
\end{figure}

The short-term frequency noise of the ECDL is dominated by current noise. We measure its linewidth (as an overlapped Allan deviation) by monitoring the PDH signal of the SHG cavity (Sec.~\ref{sec:SHG}) on timescales faster than our piezo bandwidth. The free-running linewidth is 100~kHz at 15~$\mu$s, which is consistent with the current-noise specifications of our commercial controller (Thorlabs LDC202C). We use this PDH signal to feed back on the current and narrow the linewidth by a factor of five. Fig.~\ref{fig:noise} displays the reduction of the frequency noise as a function of time. Independent measurements of the SFG cavity's PDH signal confirm that the frequency noise is indeed reduced. Narrowing the linewidth significantly reduces amplitude fluctuations in the second- and third-harmonic light.

Two tapered amplifiers (m2k Laser m2K-TA-0950-1500) boost the power available to the nonlinear conversion stages. Each amplifier produces 1.5~W of the fundamental light with of order 10~mW of input light. A combination of aspheric, cylindrical, and spherical lenses couple this light into polarization-maintaining fibers. Including loss from optical isolators and imperfect matching of the TA mode to the fiber, we typically deliver 35--40\,\% of this light to the nonlinear stages.

\section{Second harmonic generation (SHG)}\label{sec:SHG}

The second-harmonic generation (SHG) stage uses a nonlinear crystal in a power-enhancement cavity that is resonant with the fundamental. For this stage, we considered two crystals: LBO and bismuth triborate (BiBO, BiB$_3$O$_6$). The BiBO crystal benefits from a four-times larger effective nonlinear coefficient (3.36~pm/V in BiBO versus 0.81~pm/V in LBO). BiBO suffers from a four-times larger walkoff, however. The most dramatic problem with BiBO is a significant photorefractive effect at high blue powers~\cite{rusevaOC2004,jangOM2009}. While we eventually chose to run with LBO, we successfully generated blue light with BiBO in an enhancement cavity. We found that we could produce approximately 40\,\% more 470~nm light with BiBO than with LBO for the same input powers. At the highest powers, however, the photorefraction caused the output power to decline on timescales of tens of minutes and eventually induced oscillations in the crystal's refractive index (and thus in the cavity mode, cavity input coupling, and blue conversion) with periods of several seconds. Only with 470~nm output powers below 40~mW would the SHG stage produce consistent output for several hours at a time. While the LBO crystal has a lower nonlinear coefficient, it displays no photorefraction at these powers.

The 10-mm-long LBO crystal is cut for type-I conversion (ooe) with critical birefringent phase matching at room temperature ($\theta = 90^\circ$, $\phi=19.8^\circ$)~\cite{SNLO}. The index of refraction is 1.608 for both wavelengths. Its input and output facets are cut at Brewster's angle for the fundamental wavelength ($\theta_{\rm B} = 58.1^\circ$). Compared to an anti-reflection coating, the Brewster-cut facet has lower loss at the fundamental and allows for compensation of astigmatism in the cavity due to the off-axis incidence on the spherical cavity mirrors. An important disadvantage is an approximately 20\,\% reflection loss of the second harmonic, which is polarized orthogonal to the fundamental. A side of the crystal is polished, and we mount the crystal such that the reflected light is available for monitoring and diagnostic purposes.

The cavity mirrors are positioned such that the primary waist (in the crystal) is first-order insensitive to the placement of the focusing mirrors~\cite{wilsonAPB2011} and astigmatism is compensated to create a circular secondary waist. The round secondary waist simplifies mode-matching into the cavity. The curved cavity mirrors (M3, M4) have a $75$-mm radius of curvature and are placed 39~mm from the crystal facet with the full-opening angle of the incident-to-reflected beam equal to $23.6^\circ$. We calculate that the focus of the fundamental in the crystal has a horizontal waist of 48.6~$\mu$m and a vertical waist of 31.7~$\mu$m. The cavity free spectral range is measured to be 488~MHz.

Double-refraction in the crystal causes second-harmonic walkoff of $\rho=11.5$~mrad. This walkoff, combined with our crystal length $l$ and the wavenumber $k_1$ of the fundamental beam, give a crystal parameter (in the notation of Ref.~\cite{boydJAP1968}) $B\equiv \rho\,(l k_1)^{1/2}/2 = 1.9$. Both the Boyd--Kleinman analysis~\cite{boydJAP1968} and that of Freegarde, \emph{et al.}\cite{freegardeJOSAB1997}, which extends the analysis to elliptical beams, indicate that our focus is somewhat larger than optimal. It should be possible to increase the blue output at the expense of either a significantly longer cavity or an elliptical secondary waist.

The positions and angles of the mirror mounts are fixed with dowel pins in a custom-machined aluminum baseplate. One flat mirror (M2) is glued to a piezoelectric transducer, which allows active stabilization of the cavity length. This mirror is small (6.0~mm diameter, 2.3~mm thick) to minimize mass-loading the piezo. We monitor the reflected light from the cavity and lock it to resonance with the Pound--Drever--Hall technique~\cite{dreverAPB1983}. The cavity's finesse of 600 
 is high enough that the laser's modulation sidebands are nearly completely reflected when the cavity is on resonance. The ideal input-coupler (M1) has a transmission at the fundamental that matches the loss in the cavity (including conversion loss)~\cite{adamsOptComm1990}. We have tried mirrors with transmissions of 3.0\,\%, 2.2\,\%, 0.9\,\%, and 0.4\,\% and found the latter two to give comparable conversion efficiency. The data here were taken with the 0.4\,\% (99.6\,\% reflection) input coupler. Two spherical singlet lenses mode-match the collimated fiber output to the cavity.

\begin{figure}
\centering
\includegraphics[width=0.6\linewidth]{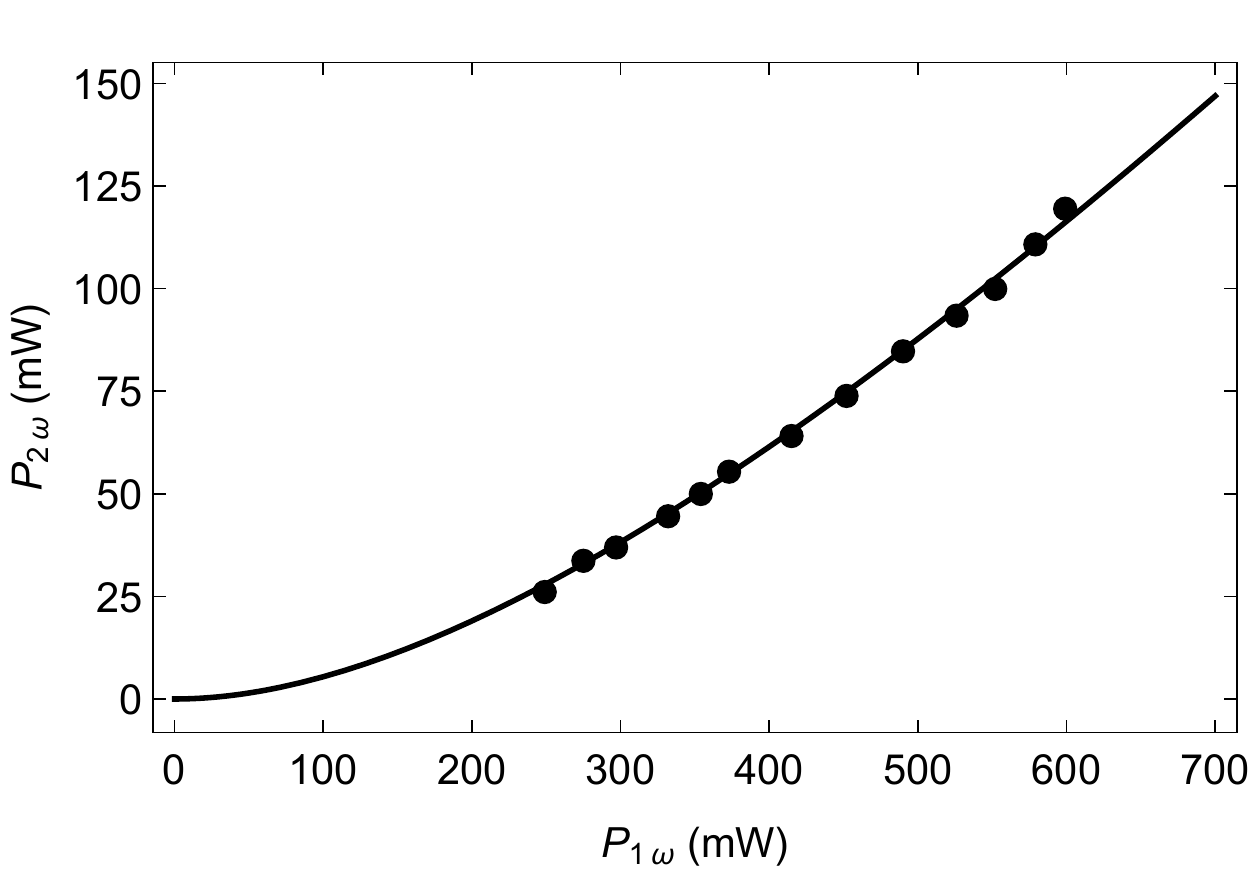}
\caption{Output power versus input power for the SHG cavity. The theory curve corresponds to nonlinear coefficient $\gamma_{\rm SHG} = 1.0\times10^{-5}~{\rm W}^{-1}$, passive cavity loss of 0.30\,\%, and includes a 20\,\% reflection loss at the crystal output facet.}
\label{fig:SHG_power}
\end{figure}

Fig.~\ref{fig:SHG_power} shows the output power of the SHG stage as a function of input power. Also plotted is a model~\cite{adamsOptComm1990} that corresponds to a second-harmonic generation coefficient of $\gamma_{\rm SHG} = P_{2\omega}/P_c^2 = 1.0\times10^{-5}~{\rm W}^{-1}$ and a passive cavity loss (excluding loss from conversion and the input coupler transmission, and stated as an equivalent reflectivity) of $R_m=0.9970$. Approximately 15\,\% of the incident fundamental light is reflected from the input coupler in a manner more consistent with a mode mismatch than an impedance mismatch. At 600~mW of IR light incident on the cavity, we measure 120~mW of blue light through mirror M4. The transmission through the input coupler is 81\,\%. The IR transmission through the high-reflectivity mirror M3 (transmission 0.04\,\%) is 47~mW, which indicates a circulating power of $P_c = 117$~W. This power-enhancement of nearly 200 agrees with that calculated from the cavity linewidth when scanning the cavity length.

\section{Sum frequency generation (SFG)}

The second harmonic light mixes with additional fundamental light in the sum-frequency generation (SFG) stage. This stage is doubly resonant to enhance the power at both wavelengths. The 10-mm-long BBO crystal is cut for type-I conversion (ooe) with critical birefringent phase matching at room temperature ($\theta = 36.0^\circ$)~\cite{SNLO}. The crystal is cut for normal incidence and anti-reflection coated at the fundamental and both harmonics. Since we want the fundamental and second harmonic to traverse nearly identical paths in the cavity, a Brewster cut will not work here. Dispersion in the BBO means the two beams have slightly different indices of refraction and thus would not be overlapped after refracting at the crystal facet. The indices are 1.657, 1.681, 1.673 for 940~nm, 470~nm, and 313~nm.

The dispersion in the BBO crystal must be compensated by other dispersive elements. We use a pair of 1.0-mm-thick fused silica plates~\cite{mesAPL2003}. In addition to dispersion compensation, the second plate compensates any displacement from the first plate. Each plate is mounted near Brewster's angle. Rotating the plates through 10~mrad provides a half-wavelength shift at 470~nm in the relative path length. Drifts in the BBO dispersion from temperature or other effects require active control of the plate angles. We mount them on counter-rotating galvanometers (Cambridge Technology model G108). It is important that the galvos be of the ``open-loop'' design. While ``closed-loop'' galvos dominate the current market because of their faster response times, this purpose does not need the increased speed and we have found their encoders to have insufficient resolution for this task.

Because of the normal-incidence crystal cut, the astigmatism from the off-axis incidence on the spherical mirrors cannot be compensated. Thus we make the cavity opening angle as small as practical -- a full-angle opening of $12^\circ$. We again position the cavity mirrors such that the primary waist is first-order insensitive to the placement of the focusing mirrors. These mirrors (M7, M8) have $75$-mm radii of curvature and are placed 37~mm from the crystal facet. The secondary waist has an ellipticity of 0.91, which does not pose a significant challenge to mode-match. The primary's waist is close to round with a calculated size of 27.3~$\mu$m for the fundamental. The second-harmonic's waist is smaller by the square root of the wavelength ratio. The cavity free spectral range is measured to be 406~MHz.

The analysis for SFG is formally equivalent to that of SHG when the parameters are appropriately redefined and the beams have the same confocal parameter~\cite{boydJAP1968}. This is the case for our beams, which share the same optical resonator. The third-harmonic walkoff is $\rho=79.1$~mrad, which corresponds to a crystal parameter $B=16$. The resulting analysis indicates that our waist is nearly optimal (9\,\% too large).

A dichroic mirror combines the fundamental and second harmonic and sends them towards the cavity. They couple in through the same mirror (M5), which is coated with a nominal 98\,\% reflectivity at both wavelengths. The fundamental light comes from a single-mode, polarization-maintaining fiber and is mode-matched to the cavity by use of two spherical singlets. The second harmonic light comes from the SHG cavity and requires more conditioning. A spherical lens collimates the SHG output beam and two cylindrical lenses correct for residual astigmatism and ellipticity. Then two spherical lenses mode-match into the SFG cavity. In principle, fewer lenses are needed, but we find the alignment easier when the function of each lens is decoupled from the others. Along the way, a half-waveplate rotates the polarization by $90^\circ$.

The cavity is kept on resonance with two servos acting on different timescales. The light reflected off the input coupler is split by a dichroic mirror and monitored on two detectors. The reflected IR light is used in a PDH lock and controls a piezo just as in the SHG cavity. With the cavity actively locked to the IR light, the galvo angles are modulated by $\approx 10~\mu$rad at 200~Hz. The signal from the reflected blue light is demodulated in a lock-in amplifier. A servo controller locks the dispersion such that the cavity is resonant in the blue as well.

\begin{figure}[tb]
\centering
{\includegraphics[width=0.75\linewidth]{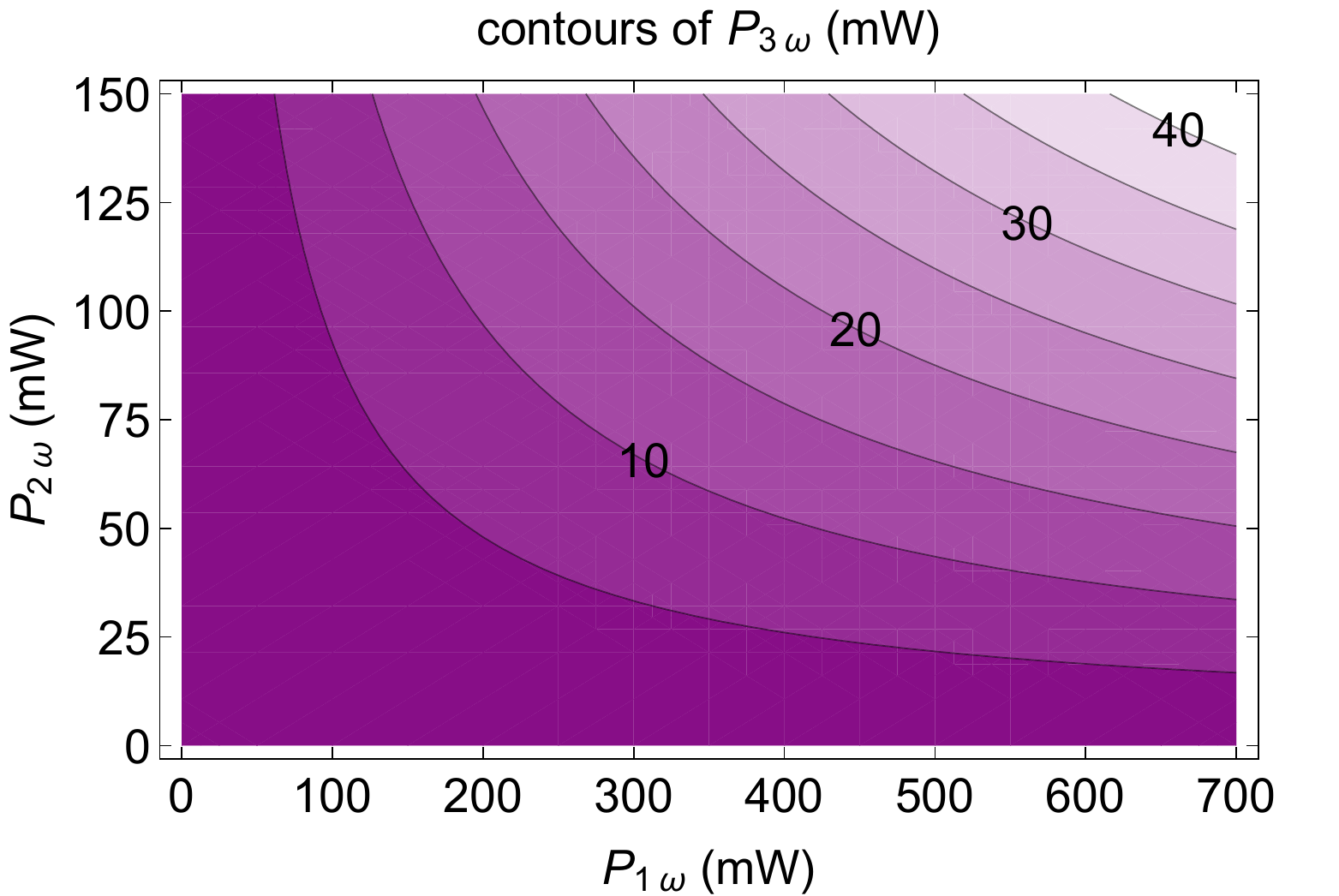}}
\caption{Model of output versus input powers for the SFG cavity. Measured output powers agree to within 15\,\% of the model. Input powers accessible with our system are up to approximately 630~mW in the fundamental ($1\omega$) and 120~mW in the second harmonic ($2\omega$). The model parameters are a nonlinear coefficient $\gamma_{\rm SFG} = 2.94\times10^{-4}~{\rm W}^{-1}$, passive cavity losses of 1.9\,\% in the fundamental and 1.8\,\% in the second harmonic, and mode-match efficiencies of 79.4\,\% in the IR and 84.7\,\% in the blue.}
\label{fig:SFG_power}
\end{figure}

To characterize the SFG conversion efficiency, we measure the UV output at several blue and IR input powers, while monitoring the circulating power and input-coupling fraction for each beam. We fit a model~\cite{kanedaAO1997} of doubly resonant SFG to these data and plot the result in fig.~\ref{fig:SFG_power}. Measured output powers agree to within 15\,\% of the model. The best-fit model parameters are a sum-frequency generation coefficient $\gamma_{\rm SFG} = P_{3\omega}/(P_{1\omega,c}P_{2\omega,c}) = 2.94\times10^{-4}~{\rm W}^{-1}$, passive losses of $R_m=0.981$ and $0.982$, and mode-match efficiencies of $0.794$ in the IR and $0.847$ in the blue. We measure 32~mW of UV output with 632~mW of IR and 105~mW of blue input. Leakage through mirror M7 (transmission $6.0\times10^{-5}$ at 940~nm and $8.1\times10^{-5}$ at 470~nm) indicate circulating powers of 27~W in the IR and 3.7~W in the blue. We have observed UV powers as high as 36~mW.

When running, the laser system uses five servos. The HeNe-stability-transfer lock is slow enough that we implement it in software. We implement the remaining four servos in two commercial FPGA-based devices (Toptica DigiLock 110), which contain two servos each. Each device controls one cavity. For the SHG cavity, one servo locks the cavity length with the piezo, while the other servo feeds back to the master laser to remove short-term frequency noise. For the SFG cavity, one servo locks the cavity length with the piezo, while the other implements the galvo dispersion lock.  In both cases, the two servos control processes on significantly different timescales. The DigiLock includes an FPGA-based lock-in amplifier for the galvo lock. The PDH locks use external mixers and amplifiers to condition the signal. The locks are interdependent and need to work simultaneously to produce UV light. It is useful for the servos to have an automated relock function to recover from sharp acoustic or vibrational disturbances. The Digilock has such a function, as do others including the open-source servo in~\cite{leibrandtRSI2015}.

\section[Use with trapped \Be~ions]{Use with trapped \boldmath{$^9$}Be\boldmath{$^+$}~ions}

We use the laser to Doppler cool trapped \Be~ions. We have used these atomic ions to sympathetically cool BeH$^+$ molecular ions. The ions are confined in a linear radiofrequency (Paul) trap~\cite{paulRMP1990}. The trap comprises machined OFHC copper electrodes mounted with titanium dowels and screws onto MACOR spacers. The radial ion--electrode distance is $r_0 = 1.2$~mm. Two opposing electrodes are electrically connected to a helical resonator~\cite{macalpineProcIRE1959}. Their electric potential is driven at $\Omega/(2\pi) = 35~{\rm MHz}$ and several hundred volts. 
 The other two electrodes are held at rf ground. They are each divided into five segments with independently controllable potentials. The center three electrodes have widths of $2 z_0 = 3.0$~mm and inter-electrode gaps of approximately 0.1~mm. Potential differences of 4~V across the central three electrodes produce a \Be~axial frequency of $\omega_z/(2\pi) = 375~{\rm kHz}$.

\begin{figure}[tb]
\centering
{\includegraphics[width=0.75\linewidth]{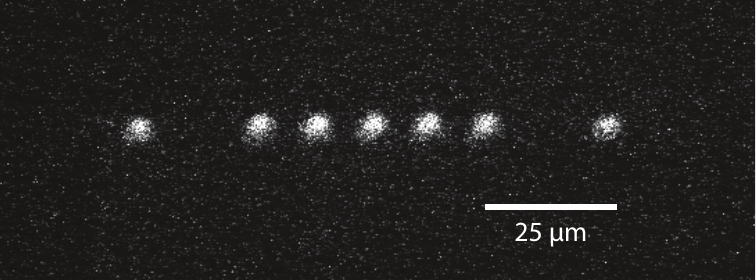}}
\caption{Fluorescence from seven trapped \Be~ions with gaps indicating two sympathetically cooled BeH$^+$ ions.}
\label{fig:ions}
\end{figure}

To load \Be, we run a current through a beryllium-wrapped tungsten wire, which creates a flux of neutral Be through the trap. An electron beam bombards the atoms, and any atom ionized within the trapping volume is trapped. We load BeH$^+$ through a chemical reaction between trapped \Be~and H$_2$ background gas. Our vacuum system has a precision leak valve to allow the controlled admission of gas, though the BeH$^+$ loaded to date have been from less-controlled reactions with residual gas in our chamber. We tune the laser system to several hundred megahertz below the \Be~$^2S_{1/2}$--$^2P_{3/2}$ transition (313.13292~nm~\cite{bollingerPRA1985}). A Glan-laser polarizer and quarter-waveplate create a pure circular polarization to address the atomic ions' cycling transition. Fig.~\ref{fig:ions} shows the fluorescence of seven \Be~ions with gaps in the chain from two BeH$^+$ ions.

\section{Conclusion}

In conclusion we have constructed a tunable UV laser near 313~nm capable of achieving 36~mW output power and based on third-harmonic-generation of an external-cavity diode laser and tapered amplifiers. We have demonstrated its use in optical-pumping, detection, and Doppler-cooling of \Be~atomic ions. This laser system produces sufficient power to drive stimulated Raman transitions~\cite{winelandJRNIST1998} in~\Be. By modulating the laser to add an optical sideband on-resonance with the \Be~transition, it should be possible to use this single laser for both coherent and dissipative control of \Be, including motional ground-state cooling and coherent sideband transitions. Such single-laser systems (based on dye or fiber lasers) have been implemented for the {Mg$^+$} ion~\cite{jostThesis2010,hemmerlingAPB2011} by use of an electro-optic modulator to add the sideband before a SHG stage. In our setup, we could use such a modulator or could modulate the current of the ECDL directly. We have been able to modulate our ECDL at frequencies as high as 13~GHz. In order for these sidebands to persist onto the UV output, the modulation frequency must be a multiple of the SFG cavity's free-spectral range. Such a diode-laser-based system will prove fruitful for work with trapped \Be~ions in quantum information processing and precision measurements.

\section*{Funding}
National Science Foundation (NSF) (CAREER PHY-1255170); Research Corporation for Science Advancement (CCSA 20929).

\section*{Acknowledgments}
The authors thank D. DeMille and J. Barry for assistance on implementing the Fabry--P\'{e}rot stabilization transfer, L. Hunter for the use of the traveling Michelson wavelength meter, and L. Weiss and A. Frenett for assistance constructing the master laser.

\end{document}